\documentstyle[12pt,epsfig]{elsart}
\begin{document}

\newcommand{\re}{\mathop{\mathrm{Re}}}
\newcommand{\im}{\mathop{\mathrm{Im}}}
\newcommand{\D}{\mathop{\mathrm{d}}}
\newcommand{\I}{\mathop{\mathrm{i}}}

\noindent {\Large DESY 03--091 \hfill ISSN 0418-9833}

\noindent {\Large July 2003}

\vspace*{1cm}

\begin{frontmatter}

\journal{}
\date{}

\title{
Two-color FEL amplifier for femtosecond-resolution pump-probe
experiments with GW-scale X-ray and optical pulses}



\author[DESY]{J.~Feldhaus},
\author[DESY]{M.~K\"orfer},
\author[DESY]{T. M\"oller},
\author[DESY]{J. Pfl\"uger},
\author[DESY]{E.L.~Saldin},
\author[DESY]{E.A.~Schneidmiller},
\author[DESY]{S.~Schreiber},
and \author[Dubna]{M.V.~Yurkov}

\address[DESY]{Deutsches Elektronen-Synchrotron (DESY),
Notkestrasse 85, Hamburg, Germany}

\address[Dubna]{Joint Institute for Nuclear Research, Dubna,
141980 Moscow Region, Russia}

\begin{abstract}

Pump-probe
experiments combining  pulses from a X-ray FEL and an optical femtosecond laser
are very attractive for sub-picosecond time-resolved studies.  Since
the synchronization between the two independent light sources to an accuracy of 100
fs is not yet solved, it is proposed to derive both femtosecond
radiation pulses from the same electron bunch but from two insertion
devices.  This eliminates the need for synchronization and developing
a tunable high power femtosecond quantum laser.  In the proposed scheme
a GW-level soft X-ray pulse is naturally
synchronized with a GW-level optical pulse, independent of any jitter in the arrival time of the electron bunches.  The
concept is based on the generation of optical radiation in a master
oscillator-power FEL amplifier (MOPA) configuration.  
X-ray radiation is generated in an X-ray undulator inserted between
the modulator and radiator sections of the optical MOPA scheme.  An
attractive feature of the FEL amplifier scheme is the absence of any
apparent limitations which could prevent operation in the femtosecond regime
in a wide (200-900 nm) wavelength range. A commercially available long (nanosecond)
pulse dye laser can be used as seed laser.

\end{abstract}

\end{frontmatter}

\clearpage

\setcounter{page}{1}

\section{Introduction}

Time-resolved experiments are used to monitor time-dependent phenomena.
In a typical pump-probe experiment a short probe pulse follows a short
pump pulse at some specified delay. The pump pulse triggers the system,
for example it heats up the sample and produces a plasma, or it starts
a photo-chemical reaction, and the probe pulse causes a signal recorded
by a conventional, slow detector which reflects the state of the sample
during the brief probing. The experiment must be repeated many times
with different delays in order to reconstruct the femtosecond dynamical
process. Femtosecond capabilities have been available for some years at
visible wavelengths. However, there is a strong interest in extending
these techniques to X-ray wavelengths because they allow to probe
directly structural changes with atomic resolution.

Recent progress in free electron laser (FEL) techniques have paved the
way for the production of GW-level, sub-100~fs, coherent X-ray pulses. This has recently been demonstrated experimentally at the TESLA Test Facility (TTF) at
DESY \cite{ttf-sat-prl}, although only at vacuum ultraviolet (VUV) wavelengths down to 80 nm. First user experiments have led to exciting
results \cite{nature}. A SASE X-ray FEL (SASE stands for self-amplified
spontaneous emission) will be commissioned as a user facility in 2004
at DESY, covering the VUV and soft X-ray range
down to 6 nm wavelength.  The unique properties of this new source are
attracting much attention in a wide science community. The
short-wavelength, GW-level radiation pulses with sub-100 fs duration
are particularly interesting for time-resolved studies of transient
structures of matter on the time scale of chemical reactions.

The straightforward approach for pump-probe experiments, the combination of the
X-ray FEL with a conventional quantum laser system, is presently being 
realized at the TESLA Test Facility at DESY \cite{pp}.  The laser
system comprises a seed pulse laser, special synchronization with the
accelerator, pulse shaping, and a pump laser together with an optical
parametric amplifier (OPA). The laser will initially cover the spectral
region between 750 nm and 900 nm and will provide a train of 200
MW-level pulses with $200$~fs pulse duration synchronized with the
FEL. The main challenges are
the development of the high power OPA and the synchronization system.
The SASE FEL can produce radiation pulses shorter than 100 fs, hence
the synchronization should be at least as good. The main uncertainty is
the time jitter of the electron bunches. The latter is produced in the
magnetic bunch compressors and is estimated to $\pm 1 {\mathrm{ps}}$
for the expected $\pm 0.1 \%$ energy jitter of the electron bunch. At
the moment it is not clear by how much this time jitter can be reduced,
and where the limits are for the electronic synchronization.

\begin{figure}[tb]
\begin{center}
\epsfig{file=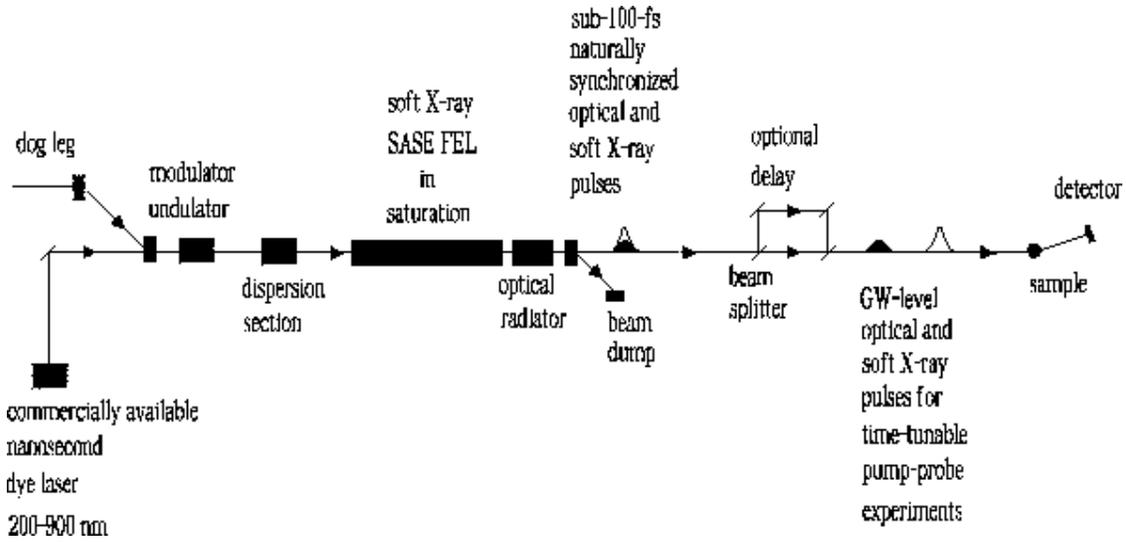,width=0.9\textwidth}
\end{center}
\caption{Scheme for pump-probe experiments employing an optical pulse
as a pump and a soft X-ray pulse as a probe or vice versa. A very long
laser pulse is used for modulation of the energy and density of the
electrons at the optical frequency.  Optical photons for pump-probe
experiments are generated by an additional insertion device (optical
radiator) using the same electron bunch }
\label{fig:pp9c}
\end{figure}

If we consider the standard technique for high-resolution time-resolved
measurements, we find that the problem of synchronization of the two
optical pulses usually does not exist at all. For very high resolution
studies with optical quantum lasers in the femtosecond regime the two
optical pulses are always derived from the same laser to ensure perfect
synchronization, and the time difference is adjusted by changing the
path length through an optical delay line.  Pump-probe experiments
combining pulses from a X-ray FEL and a quantum laser are more difficult.  The
synchronization of an independent optical laser with the FEL
pulses is the most challenging task of this type of pump-probe
technique.  Picosecond time resolved work can be performed using known
techniques.  Sub-picosecond synchronization, on the other hand, needs
further development.

The new method proposed in this paper is an attempt to get around the synchronization
obstacle by using a two step FEL process, in which two
different frequencies (colors) are generated by the same femtosecond
electron bunch. This method could be a
very interesting alternative to the "independent optical quantum laser-
SASE FEL" approach, and it has the further advantage to make a wide
frequency range accessible at high peak power and high repetition rates not so easily
available from conventional lasers.

The concept of the proposal is
schematically illustrated in Fig.~\ref{fig:pp9c}.  Two different
frequencies (colors) are generated by the same electron bunch, but in
different insertion devices.  The optical radiation  is generated in a
master oscillator-power FEL amplifier (MOPA) configuration. The X-ray radiation is generated in a X-ray undulator inserted between
modulator and radiator sections of the optical MOPA scheme. The scheme
operates as follows:  The electron beam and the optical pulse from the seed
laser enter the modulator radiator. Due to the FEL process the electron
bunch gains energy modulation at the optical frequency which is then 
transformed to a density modulation in the dispersion section.  The
density modulation exiting the modulator (i.e. the energy-modulation undulator
and the dispersion section) is about $10-20$\%. Thus, the optical seeding 
signal is imprinted in the electron bunch.
Then the electron bunch is directed to the X-ray undulator.
The process of amplification of the radiation in the main (soft X-ray)
undulator develops in the same way as in the conventional SASE FEL:
fluctuations of the electron beam current density serve as input
signal. At the chosen level of density modulation the SASE process develops
nearly in the same way as with an unmodulated electron beam because of the
large ratio of the cooperation length to the optical wavelength
\cite{sideband}. As a result, at the exit of the X-ray undulator the
electron bunch produces a GW-level X-ray pulse.  A GW-level optical pulse
is then produced when the electron bunch passes the optical radiator.  The
optical radiator is a conventional FEL amplifier seeded by the density
modulation in the electron bunch.  Although the electron beam leaving
the soft X-ray FEL has acquired some additional energy spread, it is
still a good "active medium" for an optical radiator at the end. Approximately 20\%
of density modulation is sufficient to drive the optical FEL
amplifier in the nonlinear regime and to produce GW-level optical
pulses in a short undulator.

An important feature of the proposed scheme is that the optical radiator
uses the spent electron beam.  As a result, the optical FEL can operate in saturation mode without interfering with the soft
X-ray SASE FEL operation.

We illustrate the two-color FEL amplifier scheme for the parameters of the
TESLA Test Facility at DESY. The proposed pump-probe
facility has unique features:  Both, X-ray and optical pulses, have very
high peak power in the GW range. Both wavelengths are continuously and
independently tunable in a wide range:  $200-900$~nm for the optical
pulses, and $6-120$~nm for the X-ray pulses. Both pulses have diffraction
limited angular divergence.  The spectral width of the optical pulse is
transform limited. Finally, the optical and X-ray pulses are precisely
synchronized at a femtosecond level, since they both are produced by
the same electron pulse, and there are no reasons for any
time jitter between them.

\section{Two-color FEL amplifier at the TESLA Test Facility}

\begin{figure}[b]
\begin{center}
\epsfig{file=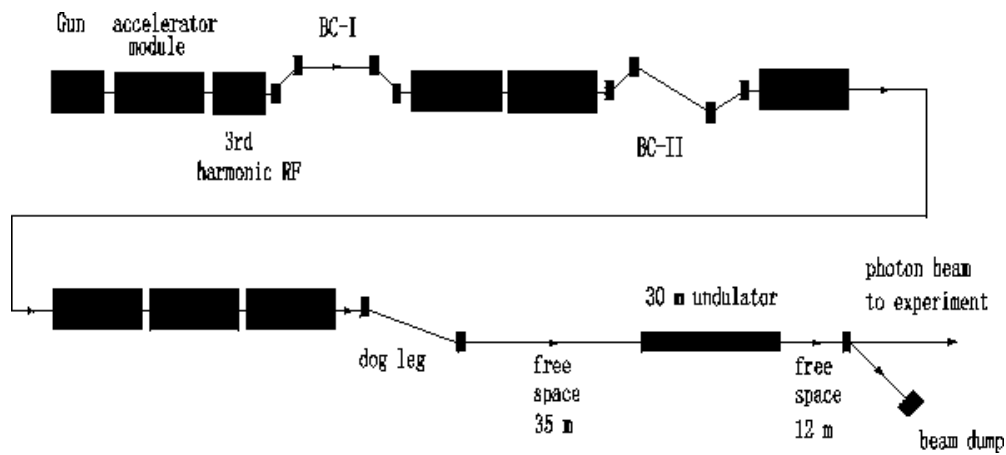,width=0.8\textwidth}
\end{center}
\caption{ Schematic layout of the soft X-ray SASE FEL facility
}
\label{fig:pp16a}
\end{figure}

Figure~\ref{fig:pp16a} shows the layout of the soft X-ray FEL in the
first phase, including the linear accelerator with two magnetic bunch
compressors and six undulator modules at the end of the accelerator
tunnel.  The first goal is to reach saturation in the soft X-ray range
with this configuration, using the six undulator modules in SASE mode.
In a second step the free space in front of the undulator will be used
to build a fully coherent soft X-ray facility based on the two stage
self-seeding concept \cite{seeding-option}.

\begin{figure}[tb]
\vspace*{-0.5cm}
\begin{center}
\epsfig{file=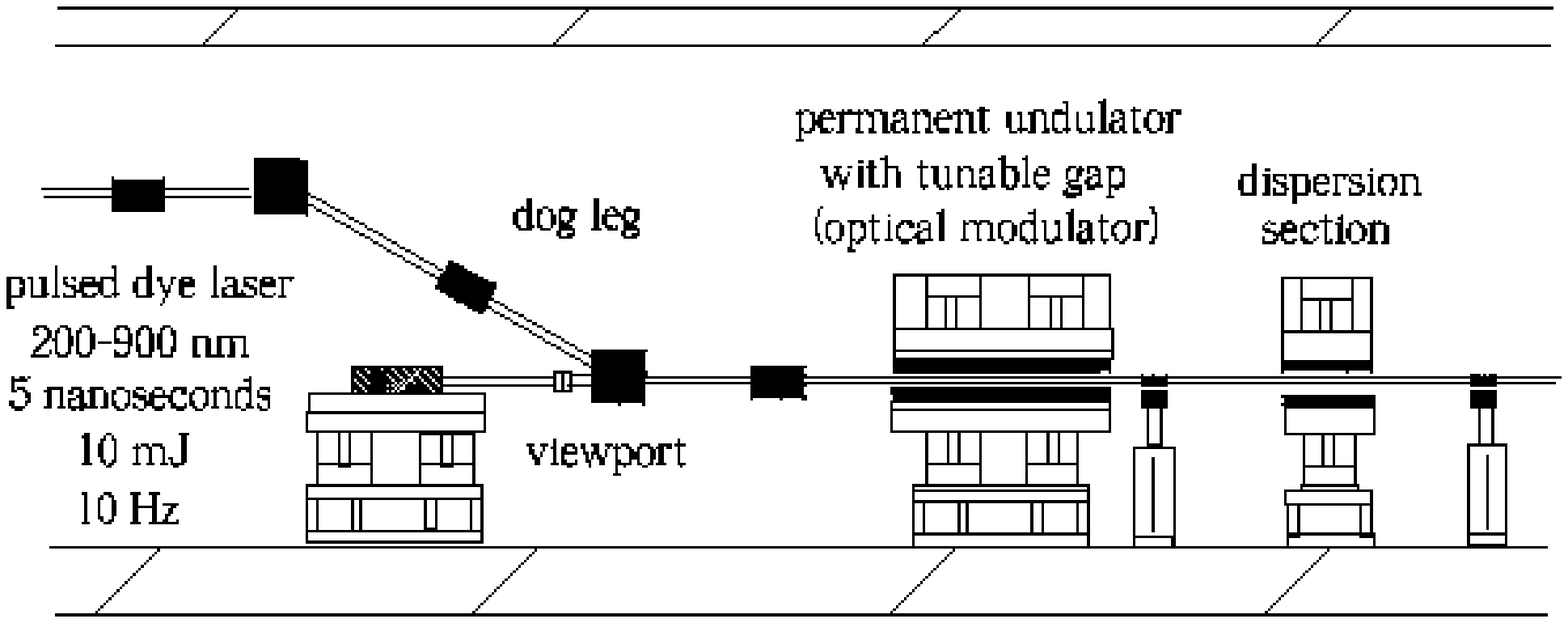,width=0.75\textwidth}

\vspace*{-1.7cm}

\epsfig{file=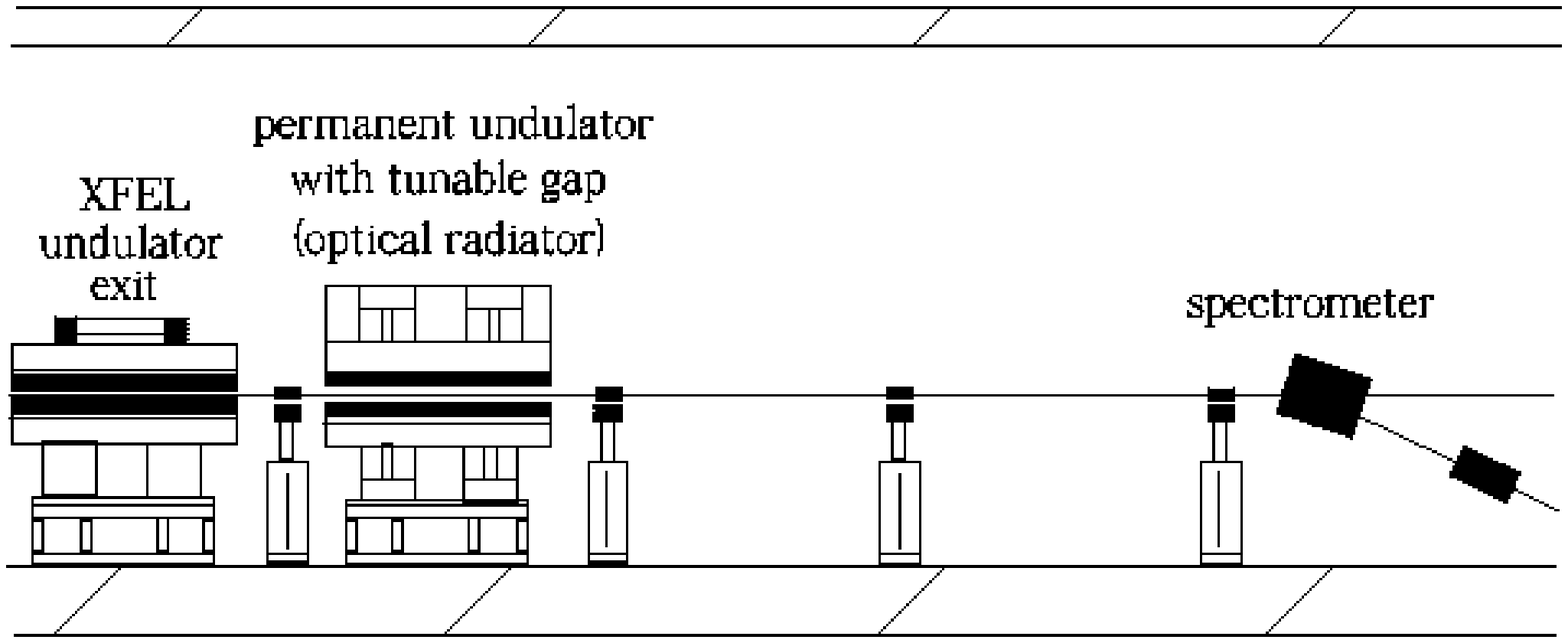,width=0.75\textwidth}
\end{center}
\caption{ Side view of the electron beam transport system, showing the
location of the seed laser and the modulator (top) and
of the optical radiator (bottom)
}
\label{fig:pp18b}
\end{figure}

\begin{table}[tb]
\caption{Parameters of the components required for the optical part of
the femtosecond pump-probe facility
}
\medskip
\begin{tabular}{l c c c }
\hline
\underline{Seed laser} \\
\hspace*{10pt} Type                             & dye \\
\hspace*{10pt} Wavelength [nm]                  & 200--900 \\
\hspace*{10pt} Pulse duration [ns]              & 5--10 \\
\hspace*{10pt} Pulse energy [mJ]                & 2--10 \\
\hspace*{10pt} Repetition rate [Hz]             & 10 \\
\underline{Undulator}\\
\hspace*{10pt} Number of modules                & 2 \\
\hspace*{10pt} Type                             & planar \\
\hspace*{10pt} Period [cm]                      & 8.2 \\
\hspace*{10pt} Gap [mm]                         & 12--30 \\
\hspace*{10pt} Peak field [T]                   & 0.5--1.8 \\
\hspace*{10pt} K-value                          & 5--14 \\
\hspace*{10pt} Number of periods                & 55 \\
\hspace*{10pt} Length of each module [m]        & 4.5 \\
\underline{Modulator chicane} \\
\hspace*{10pt} Number of dipoles magnets        & 4 \\
\hspace*{10pt} Length of each dipole magnet [m] & 0.25 \\
\hspace*{10pt} Maximum magnetic field [T]       & 0.2 \\
\hspace*{10pt} Total length of chicane [m]      & 1.5 \\
\hline
\end{tabular}
\label{tab:optical-hardware}
\end{table}

The two-color FEL amplifier employs additional undulators to generate the
second color radiation pulses. It requires free space in front of and
behind the main soft X-ray FEL undulator which has already been
foreseen at the design stage of the TTF FEL, Phase 2.  Figure
\ref{fig:pp18b} shows the location of additional hardware components:
the seed laser, the modulator undulator, a dispersion section, the main
soft X-ray SASE undulator, and the optical undulator.  The parameters
of the hardware components required for the optical part of the
pump-probe facility are listed in Table~\ref{tab:optical-hardware}.

A commercially available dye laser, for example, can be used as a seed
laser. The typical pulse energy of a dye laser system is in the range
of 2 to 10~mJ with a pulse duration of 5--10~ns, and the peak power is
in the range of 1 MW which gives us sufficient safety margin for
operation of the modulator.  The installation of a seed laser is
greatly facilitated by the fact that the magnetic chicane of the electron beam collimation system allows to insert a view port for
the input optical system. Also it is very important, that there is 
free space downstream of the main undulator (see Fig. \ref{fig:pp18b})
available for the optical radiator.

Both modulator and radiator undulators are identical tunable-gap
devices similar to those used at DORIS. The dispersion section is
composed of four standard bending magnets similar to those used at the
HERA storage ring.  Therefore, this optical radiation source could be realized at the TESLA Test Facility rather quickly
and with minimum cost expenses.

\section{Generation of two-color radiation pulses}

The principle of two-color femtosecond pulse generation has
been sketched briefly in Sections 1 and 2. Here we present a detailed
description of the physical processes in both the optical and X-ray
undulators. The parameters of the optical part of the facility have
been optimized using the time-dependent FEL simulation code FAST
\cite{fast}.  Starting point for the optimization were the parameters of the
TESLA Test Facility, Phase 2, given in an update \cite{ttf-phase2} to the
original design report \cite{ttf-cdr}.
The FEL at DESY will cover the wavelength range from 120 to
6~nm. Two modes of operation are currently foreseen:

\begin{enumerate}

\item femtosecond mode (for $\lambda = 30 - 120$~nm, pulse duration
50-100~fs);

\item short wavelength mode (for $\lambda = 6 - 30$~nm, pulse
duration 200~fs).

\end{enumerate}

The analysis of experimental results obtained at the TTF FEL, Phase 1,
showed that the local energy spread is less than 0.2~MeV
\cite{piot-simulations-ttf2} which is significantly less than the
previous project value of 1~MeV.  This low value of the local energy
spread improves significantly the operation of both the optical and the
X-ray FEL and extends the safety margin for operation.

The operation of the two-color femtosecond facility is illustrated for
the two modes of FEL operation: short wavelength mode with a bunch
shape close to gaussian, and femtosecond mode when the electron bunch
has a strongly non-gaussian shape with an intense leading peak
\cite{ttf-sat-prl,ttf-phase2}. We demonstrate that the two-color
facility will work effectively in both cases.

\subsection{Operation of the optical modulator}

   The optical modulator consists of three elements: the optical seed
laser, the modulator undulator, and the dispersion section.  The seed
laser pulse interacts with the electron beam in the modulator undulator
which is resonant with the laser frequency $\omega_{\mathrm{opt}}$, and
produces an energy modulation of $P_0$ in the electron bunch. The
amplitude of the induced beam modulation is small, typically of about a
percent only. The electron beam then passes through the dispersion
section where the energy modulation is converted to a density
modulation at the optical wavelength.  Optimum parameters of the
dispersion section can be calculated in the following way. The phase
space distribution of the particles in the first FEL amplifier is
described in terms of the distribution function $f(P,\psi)$ written in
"energy-phase" variables $P = E - E_{0}$ and $\psi = k_{\mathrm{w}}z +
\omega_{\mathrm{opt}}(z/c-t)$, where $E_{0}$ is the nominal energy of
the particle, $k_{\mathrm{w}} = 2\pi/\lambda_{\mathrm{w}}$ is the
undulator wavenumber, and $\omega_{\mathrm{opt}}$ is the frequency of
the seed radiation. Before entering the first undulator, the electron
distribution is assumed to be Gaussian in energy and uniform in phase
$\psi$:

\begin{displaymath}
f_{0}(P,\psi) = \frac{1}{\sqrt{2\pi\langle(\Delta
E)^{2}\rangle}} \exp\left(-\frac{P^{2}}{2\langle(\Delta
E)^{2}\rangle}\right) \ .
\end{displaymath}

\noindent At the exit of the first undulator the amplitude modulation
is very small, and there is an energy modulation $P_{0}\sin \psi$ only.
Thus, the distribution function at the entrance to the dispersion
section is $f_{0}(P + P_{0}\sin \psi )$.  After passing through the
dispersion section with a dispersion strength $\D\psi/\D P$, the
electrons of phase $\psi$ and energy deviation $P$ will come to a new
phase $\psi + P\D\psi/\D P$. Hence, the distribution function becomes

\begin{displaymath}
f(P,\psi) = f_{0}\left(P + P_{0}\sin\left(\psi - P\frac{\D\psi}{\D
P}\right)\right) \ .
\end{displaymath}

\noindent The integration of this distribution over energy provides the
beam density distribution, and the Fourier expansion of this function 
gives the harmonic components of the density modulation converted from
the energy modulation \cite{czonka}:

\begin{displaymath}
I/I_0 =
1 + 2 \sum^{\infty}_{n=1} \exp\left[ -
\frac{n^2}{2}\langle(\Delta
E)^{2}\rangle\left(\frac{\D\psi}{\D P}\right)^{2}\right]
J_{n}\left(nP_{0}\frac{\D \psi}{\D
P}\right)\cos(n\psi) \ .
\label{eq:h}
\end{displaymath}

\noindent Here $J_n$ is the Bessel function of n-th order.
\noindent Assuming the argument of the Bessel function to be small, we find
that maximum bunching at the fundamental harmonic
$(a_1)_{\max } =
P_{0}
/\sqrt{2.72
\langle(\Delta
E)^{2}\rangle}$
is achieved at $(\D
\psi/\D P)_{\max } = 1/{\sqrt{\langle(\Delta E)^{2}\rangle}}$.

During the passage through a long main SASE undulator the electron
density modulation at the optical wavelength can be suppressed by the
longitudinal velocity spread in the electron beam. For effective
operation of the optical FEL amplifier the value of the suppression
factor should be close to unity. A calculation 
shows that this should not be a serious limitation in the TTF case.

\subsection{Femtosecond mode operation}

\begin{figure}[b]

\begin{center}
\epsfig{file=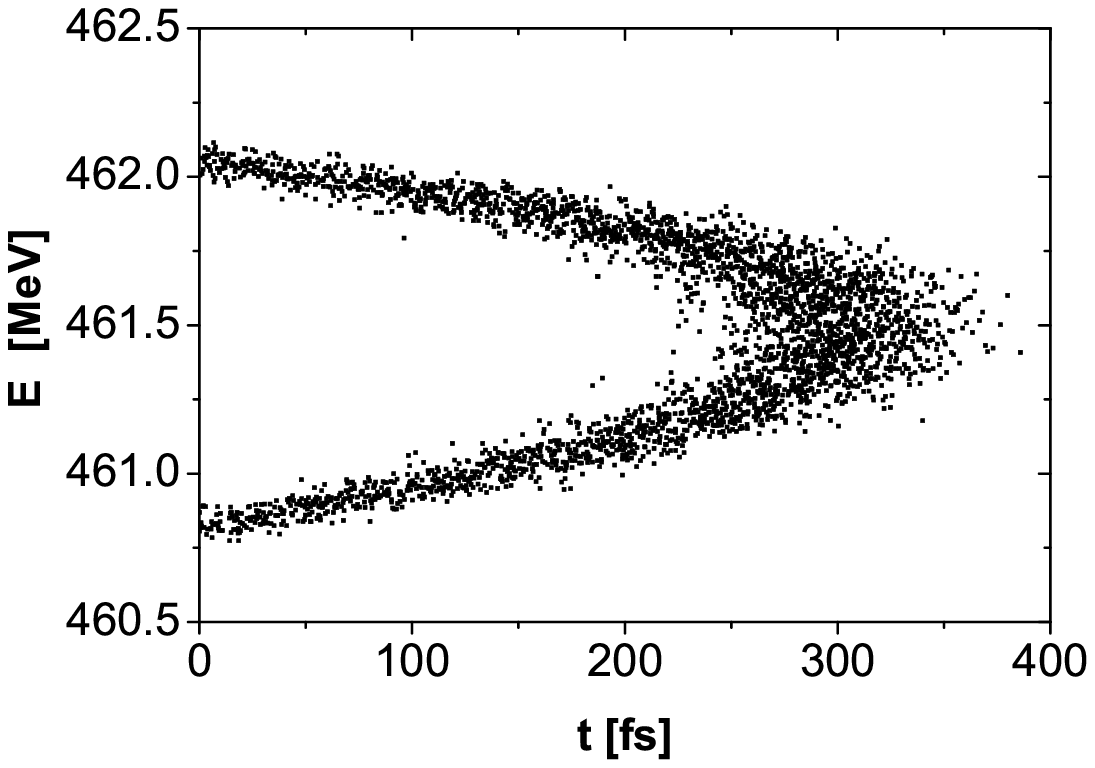,width=0.6\textwidth}
\end{center}

\caption{
Phase space distribution of electrons in the femtosecond mode after full
compression with the first bunch compressor.
The head of the bunch is on the
right. The charge of the bunch is 1 nC,
the local energy spread is 5~keV at the
entrance of the bunch compressor.
}
\label{fig:phsp-fem}


\epsfig{file=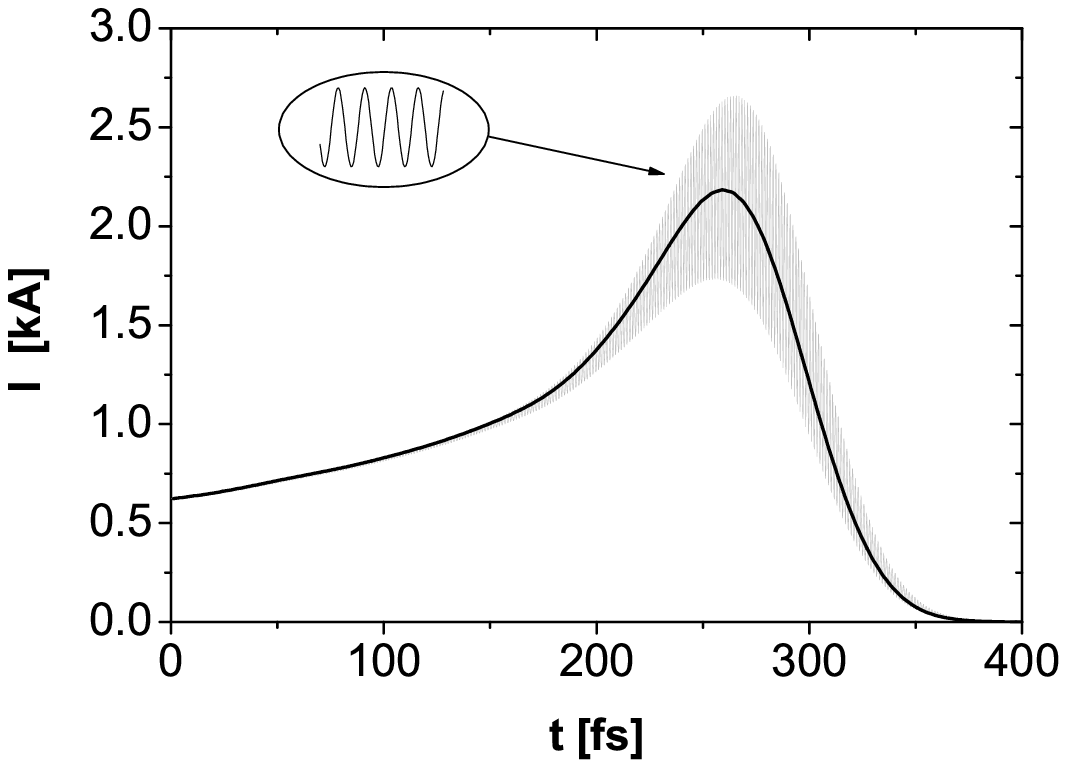,width=0.5\textwidth}

\vspace*{-63mm}

\hspace*{0.5\textwidth}
\epsfig{file=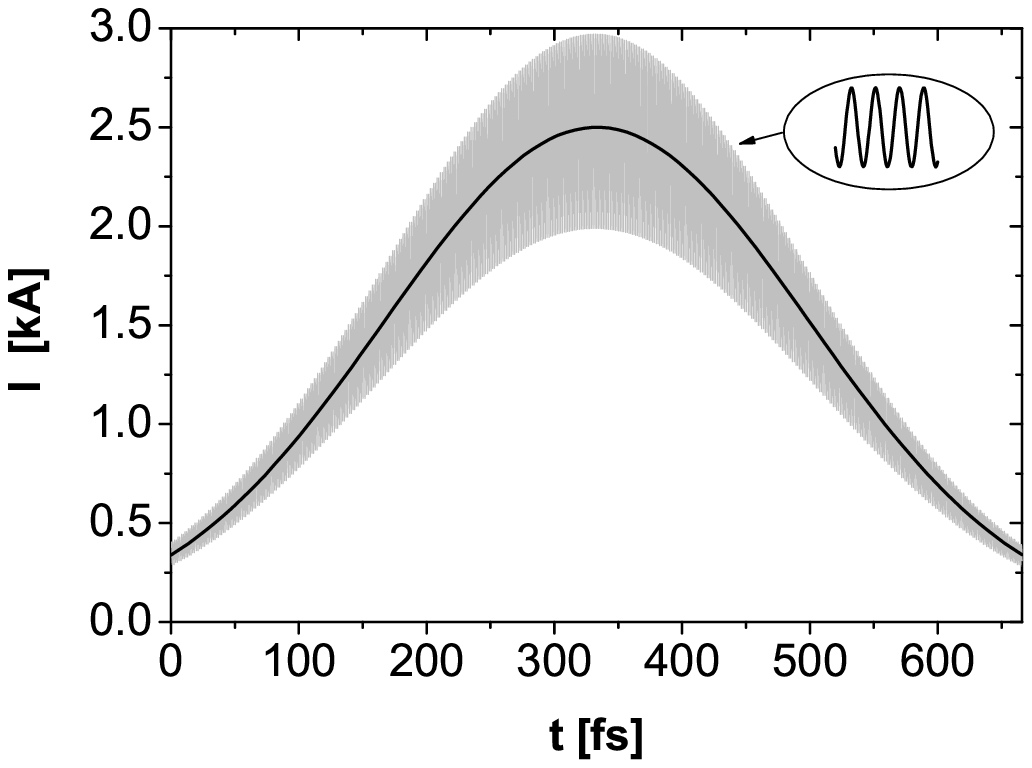,width=0.5\textwidth}

\caption{Current distribution along the bunch after the dispersion section
for femtosecond mode operation (left plot) and
for short wavelength mode operation (right plot).
The bunch is modulated with a period equal to the optical wavelength.
The solid lines show the bunch profiles at the
entrance to the optical modulator.
}
\label{fig:mod-fem}
\end{figure}

The femtosecond mode operation is based on the experience obtained
during the test runs of the TTF FEL, Phase 1 \cite{ttf-sat-prl} and it
requires one bunch compressor only. An electron bunch with a sharp
spike at the head is prepared, with an rms width of about 20 $\mu $m
and a peak current of about 2 kA. This spike in the bunch generates FEL
pulses with a duration below one hundred femtoseconds. An example of
the longitudinal phase-space distribution for a compressed beam
including the effect of RF curvature is shown in Fig.
\ref{fig:phsp-fem}, where the longitudinal bunch charge distribution
involves concentration of charges in a small fraction of the bunch
length. The longitudinal bunch profile is shown as solid line in
Fig.~\ref{fig:mod-fem}. In the femtosecond mode only the first magnetic
chicane BC-I will be active, and this will be the default mode of
operation for some time until the 3rd harmonic cavity has been
installed in the injector (see Fig. \ref{fig:pp16a}).

\begin{figure}[tb]

\epsfig{file=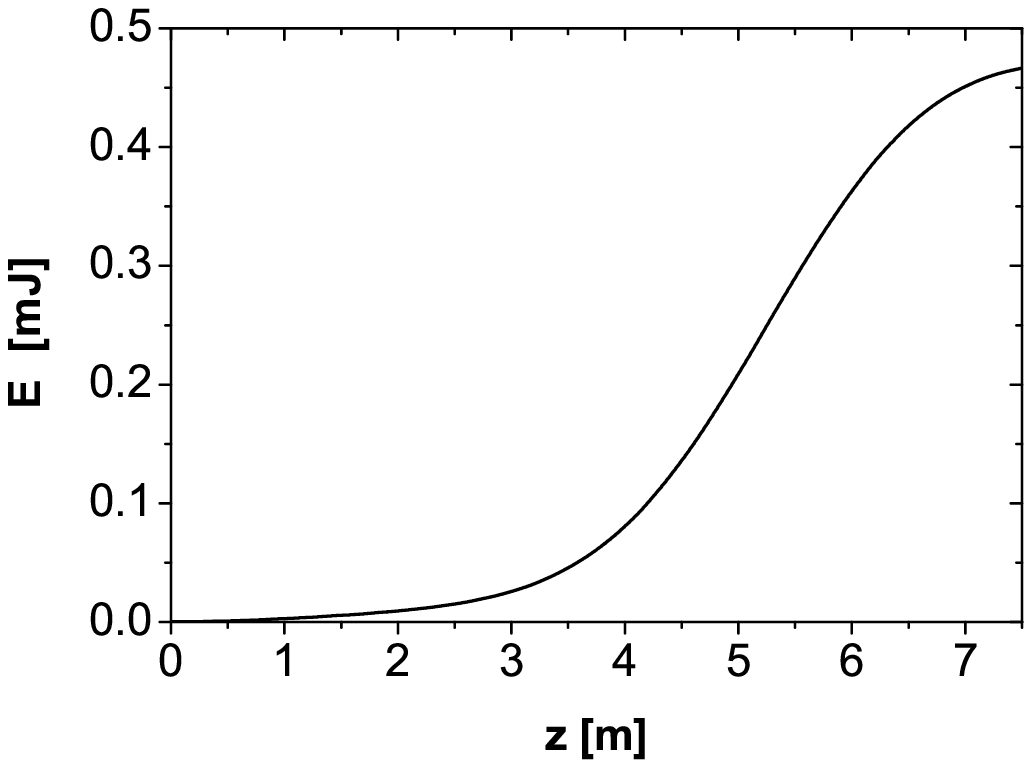,width=0.5\textwidth}

\vspace*{-63mm}

\hspace*{0.5\textwidth}
\epsfig{file=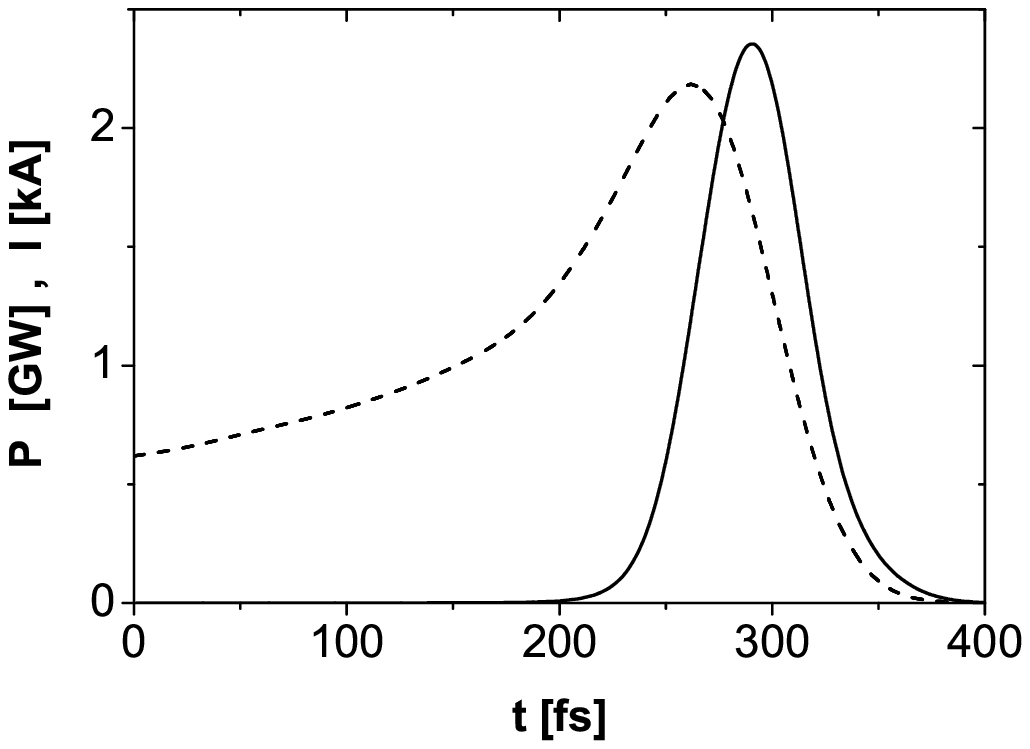,width=0.5\textwidth}

\caption{
Radiation energy in the optical pulse versus undulator length
(left plot) and time structure of the radiation pulse at the exit of
the optical radiator (right plot) in the femtosecond mode for a radiation
wavelength of 400~nm.
The dashed line indicates the profile of the electron bunch.
}
\label{fig:r400-fem}


\epsfig{file=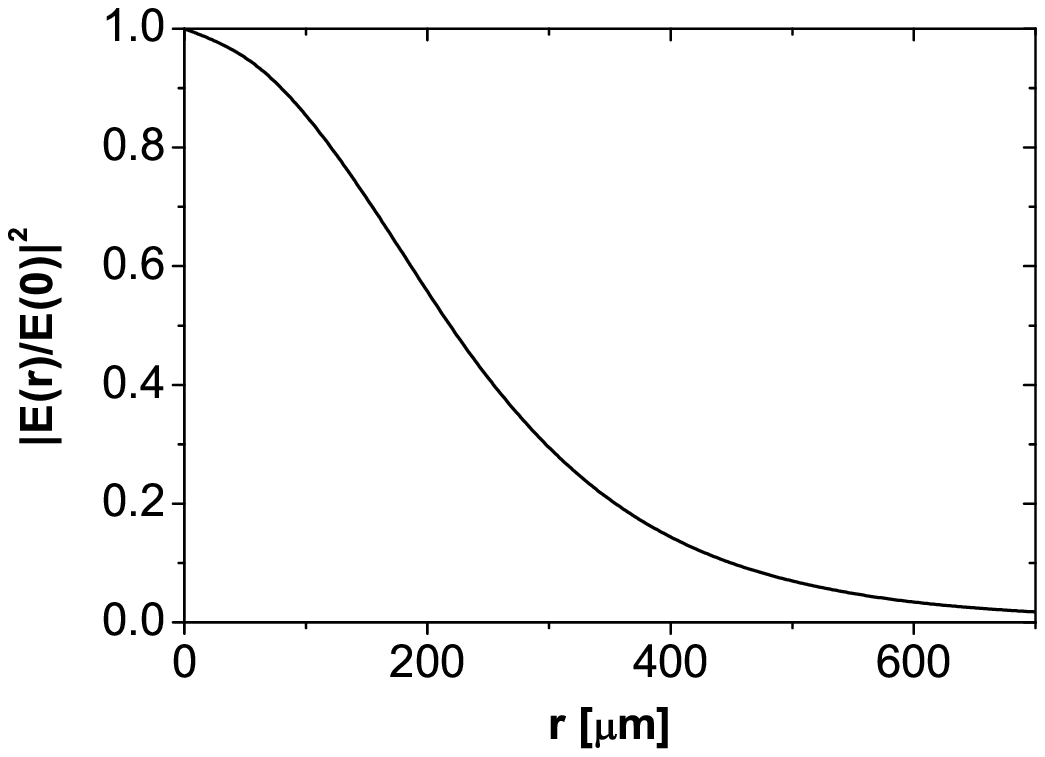,width=0.5\textwidth}

\vspace*{-63mm}

\hspace*{0.5\textwidth}
\epsfig{file=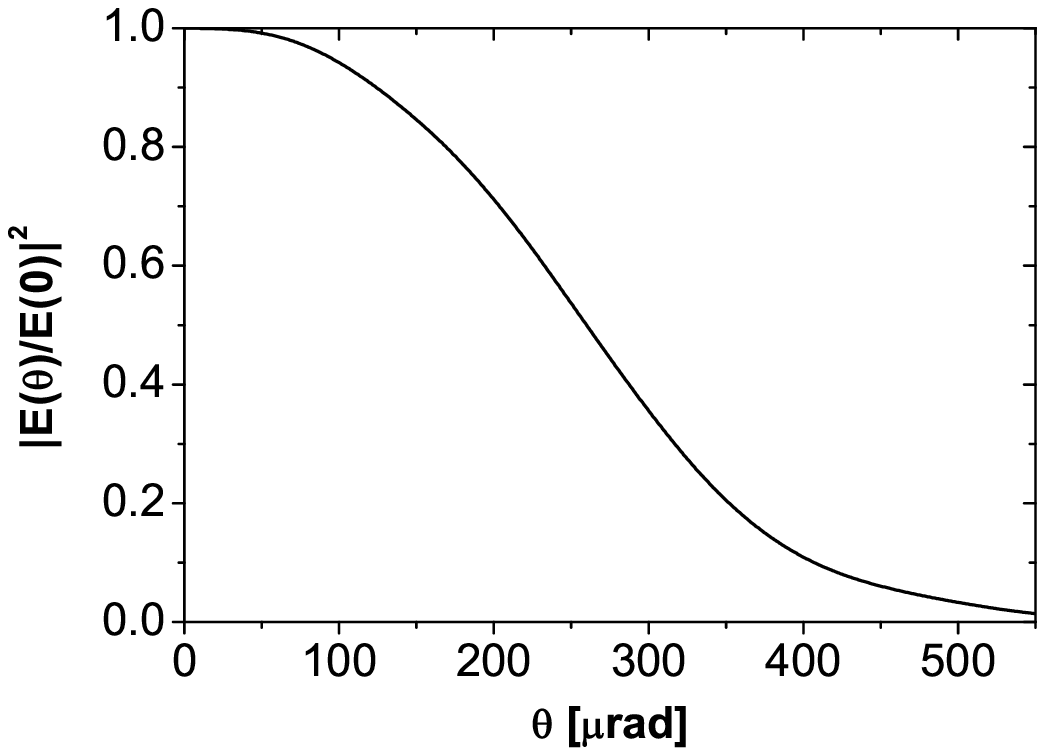,width=0.5\textwidth}

\caption{
Radial distribution of the radiation intensity at the undulator exit
(left plot) and intensity distribution in the far zone (right plot) for
400~nm wavelength in the femtosecond mode.
}
\label{fig:nf-fem-400}
\end{figure}

\begin{figure}[p]
\begin{center}
\epsfig{file=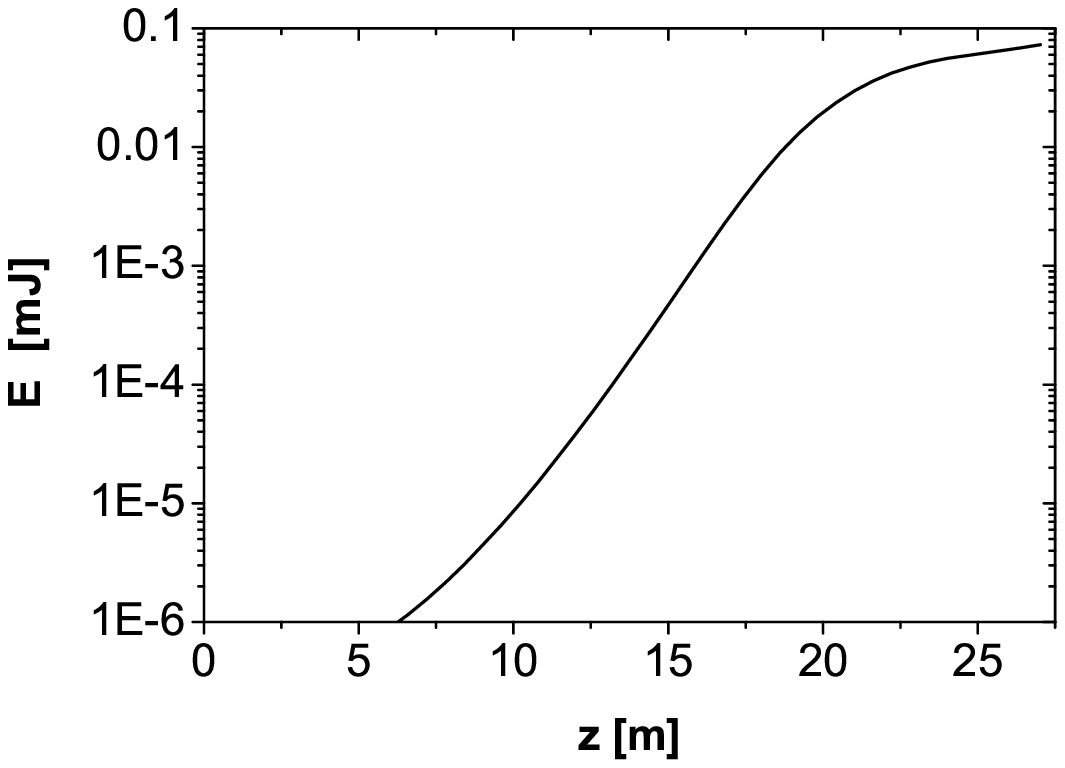,width=0.6\textwidth}
\end{center}
\caption{
Radiation energy in the X-ray SASE pulse versus undulator length for the
femtosecond mode of operation.
The radiation wavelength is 30~nm.
}
\label{fig:pz30-fem}


\epsfig{file=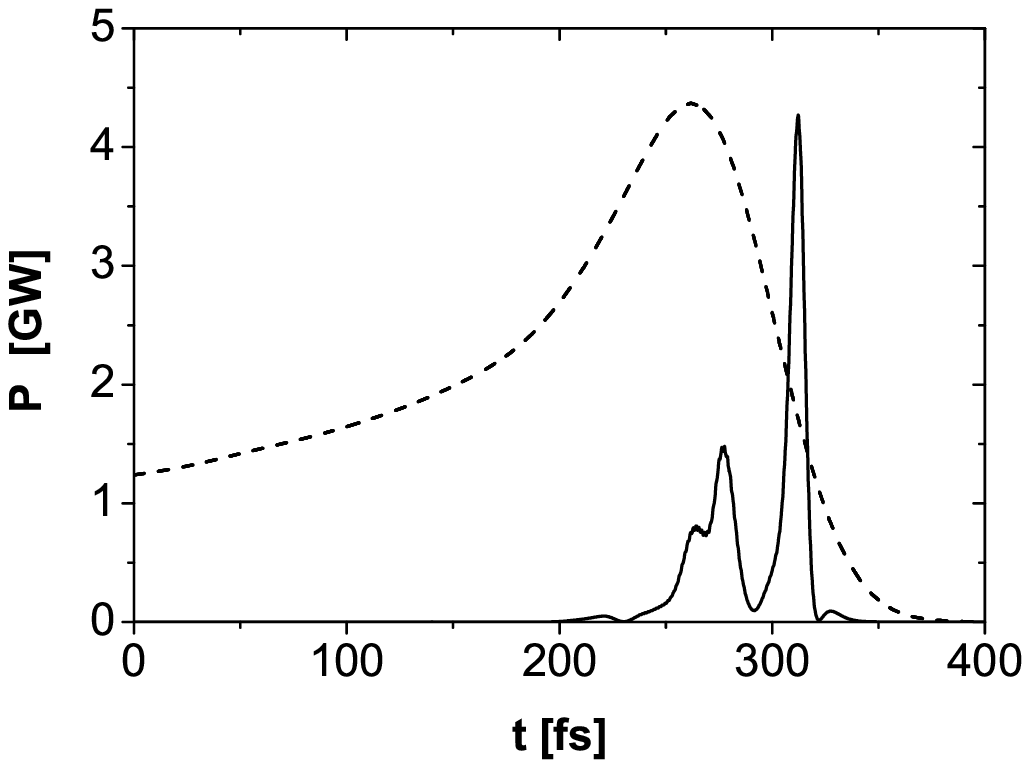,width=0.5\textwidth}

\vspace*{-63mm}

\hspace*{0.5\textwidth}
\epsfig{file=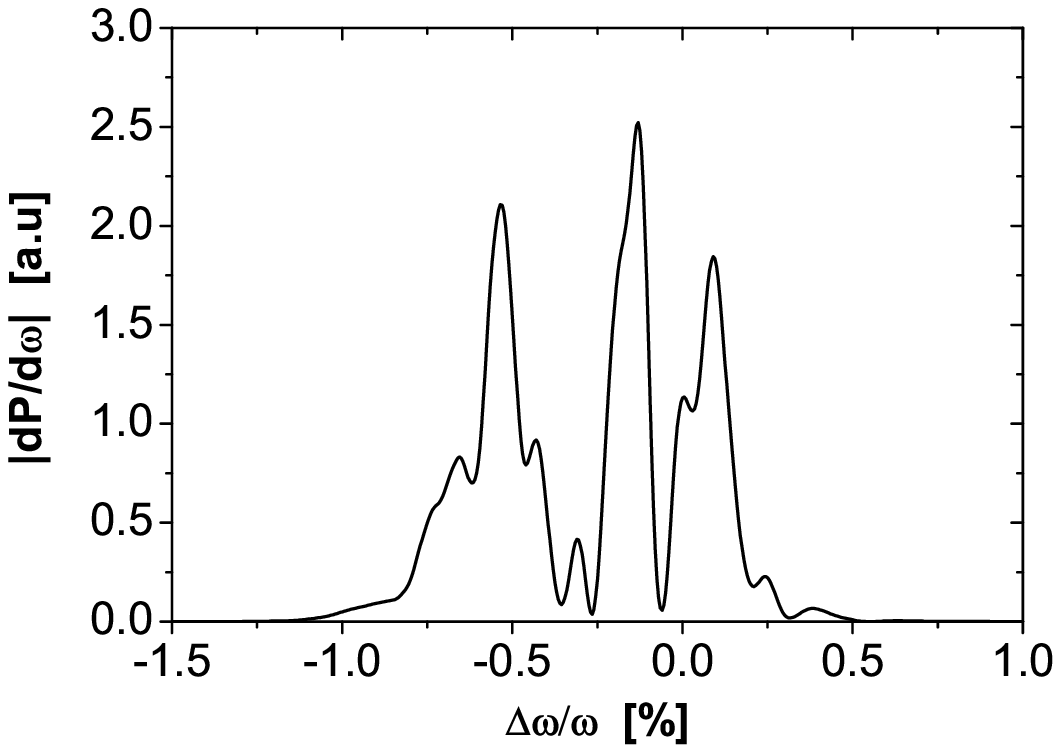,width=0.5\textwidth}

\caption{
Time (left plot) and spectral (right plot) structure of the X-ray
SASE pulse at the exit of the undulator for the femtosecond mode of
operation at a  radiation wavelength of
30~nm.
The dashed line shows the bunch profile.
}
\label{fig:ts-fem}


\epsfig{file=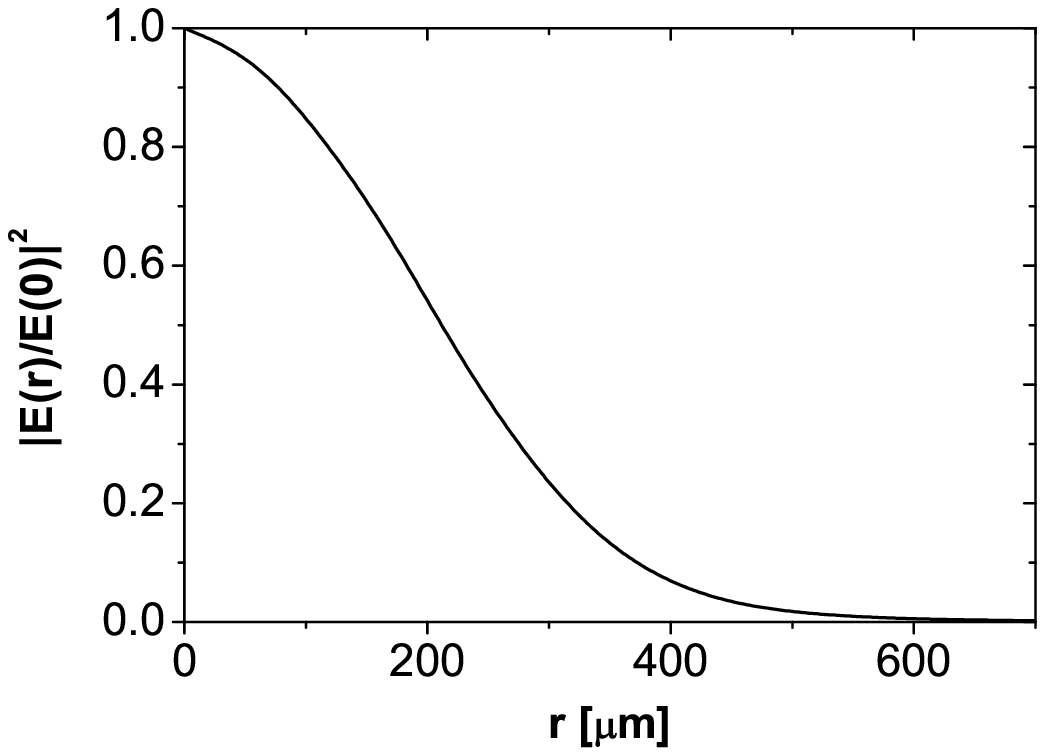,width=0.5\textwidth}

\vspace*{-63mm}

\hspace*{0.5\textwidth}
\epsfig{file=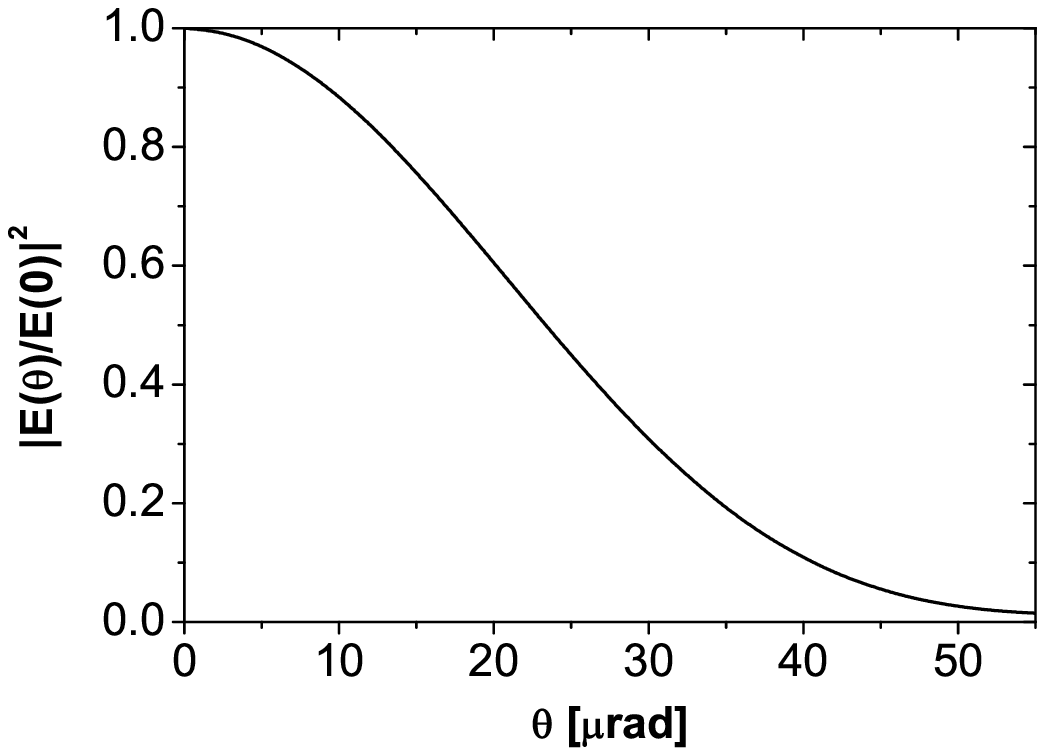,width=0.5\textwidth}

\caption{
Radial distribution of the X-ray SASE pulse intensity at the undulator exit
(left plot) and intensity distribution in the far zone (right plot) for the
femtosecond mode of operation at a
radiation wavelength of 30~nm.
}
\label{fig:nf-fem-30}
\end{figure}

The performance of the two-color facility is illustrated for an optical
wavelength of 400~nm and an X-ray wavelength of 30~nm. The bunch
profile at the entrance of the undulator is shown as a solid line in
Fig.~\ref{fig:mod-fem}.  Simulations show that the value of the slice
normalized emittance in the leading spike is about 7$\pi $~mm-mrad. An
important feature is a pronounced decrease of the local energy spread
in the head of the bunch which can be derived from the phase space
distribution (see Fig.~\ref{fig:phsp-fem}).  It will be
shown below that the combination of high peak current and low energy
spread in the leading spike will result in significant shortening of
both optical and X-ray radiation pulses.

The seed laser pulse interacts with the electron beam in the modulator
undulator and produces an energy modulation of about 100~keV in the
electron bunch. The amplitude of the induced beam modulation is less
than a percent. Then the electron beam passes through the dispersion
section where the energy modulation is converted to a density
modulation at the optical wavelength.  The grey line in
Fig.~\ref{fig:mod-fem} shows the bunch profile after the dispersion
section. The density modulation reaches an amplitude of about 20\%. It
should be noted that the beam modulation is strongly non-uniform along
the bunch. This is a consequence of the strongly varying energy spread
along the bunch, since the bunching depends on the ratio of the energy
modulation to the local energy spread. In our example the dispersion
section is tuned to obtain maximum bunching in the top of the spike.
Thus, the seeding is strongly suppressed in the tail of the bunch.

Upon leaving the dispersion section, the electron beam passes the X-ray
undulator where it produces X-ray pulses (this process will be
described below).  The electron bunch leaving the X-ray undulator has a
large induced energy spread of about 1~MeV but is still ``cold'' enough
for the generation of optical radiation. Since the bunch is strongly
modulated at the optical wavelength $\lambda_{\mathrm{opt}}$, it
readily starts to produce powerful optical radiation when it enters the
optical radiator resonant at $\lambda_{\mathrm{opt}}$.  The evolution
of the radiation energy in the optical radiator is presented in the
left plot of Fig.~\ref{fig:r400-fem}.  The right plot shows the
temporal structure of the optical pulse at the exit of the optical
radiator (at z=4.5~m).  The dashed curve in this plot presents the
electron bunch shape.  It is evident that the optical pulse is much
shorter than the electron bunch.  This is the result of the three
factors mentioned above:  the strongly non-uniform pulse profile, the
decrease of the energy spread in the head of the bunch, and the
nonuniform seeding modulation. All these factors lead to a very strong
suppression of the lasing properties of the bunch tail.  The optical
pulse has about 2 GW peak power, 50~fs FWHM pulse width, and about
100~$\mu $J pulse energy.  The installation of one additional optical
undulator would allow to increase the pulse energy to 500~$\mu $J as it
is seen from the left plot in Fig.~\ref{fig:r400-fem}.  It should be
noted that the optical pulse is completely coherent, and its spectral
width is transform-limited.  The transverse shape of the radiation
pulse at the undulator exit and its intensity distribution in the far
zone are shown in Fig.~\ref{fig:nf-fem-400}.

So far we have considered the chain for producing femtosecond optical
pulses. Let us now turn our attention to the SASE process in the X-ray
undulator. Although the electron bunch density is strongly modulated
(see Fig.~\ref{fig:mod-fem}), the SASE FEL process in the X-ray
undulator remains almost the same as for an unmodulated electron beam.
This is due to the fact that the cooperation length in the X-ray
undulator is much longer then the modulation period at the optical
wavelength. As a result, averaging of the process takes place. Of
course, the output radiation has contents of the sideband harmonic
\cite{sideband}, but its contribution to the total radiation energy is
tiny, of the order of 10$^{-4}$. Figure~{\ref{fig:pz30-fem} shows the
evolution of the energy in the X-ray pulse along the X-ray undulator,
and Fig.~\ref{fig:ts-fem} shows the time and spectral structure of the
radiation pulse at the undulator exit. The properties of the radiation
pulse are the same as for an unmodulated electron beam
\cite{ttf-phase2}.  The peak radiation power is in the GW range and the
FWHM pulse width is about 50~fs.  The transverse shape of the radiation
pulse at the undulator exit and its intensity distribution in the far
zone are shown in Fig.~\ref{fig:nf-fem-30}.

These results demonstrate that the two-color scheme for the femtosecond
mode of operation is capable to produce 50~fs long, GW-level optical
and X-ray pulses which are precisely synchronized at a femtosecond
level, since they both are produced by the same electron bunch, and
there are no effects which could affect the synchronization.

\subsection{Short wavelength mode operation}

Operation of the FEL in the short wavelength mode requires the complete
chain of the bunch compression scheme (BC-I, BC-II, and the 3rd
harmonic RF section of the injector, see Fig. \ref{fig:pp16a}).  The
operating range of the accelerator is from 460 to 1000~MeV electron
energy. The longitudinal profile of the electron bunch is shown in
Fig.~\ref{fig:mod-fem}. We illustrate the proposed two-color FEL scheme
for an X-ray wavelength of 6~nm (i.e. the minimum project value) and an
optical wavelength of 400~nm (in the middle of the tuning range).  The
parameters of the electron beam are: 1~GeV energy, 2500~A peak current,
and 2$\pi $~mm-mrad rms normalized emittance. Recent start-to-end
simulations predict a local energy spread less than $0.2$~MeV
\cite{piot-simulations-ttf2}.  In the present example we use a value of
0.5~MeV for the local energy spread in order to demonstrate that  the
proposed pump-probe scheme has sufficient safety margin.

\begin{figure}[t]

\epsfig{file=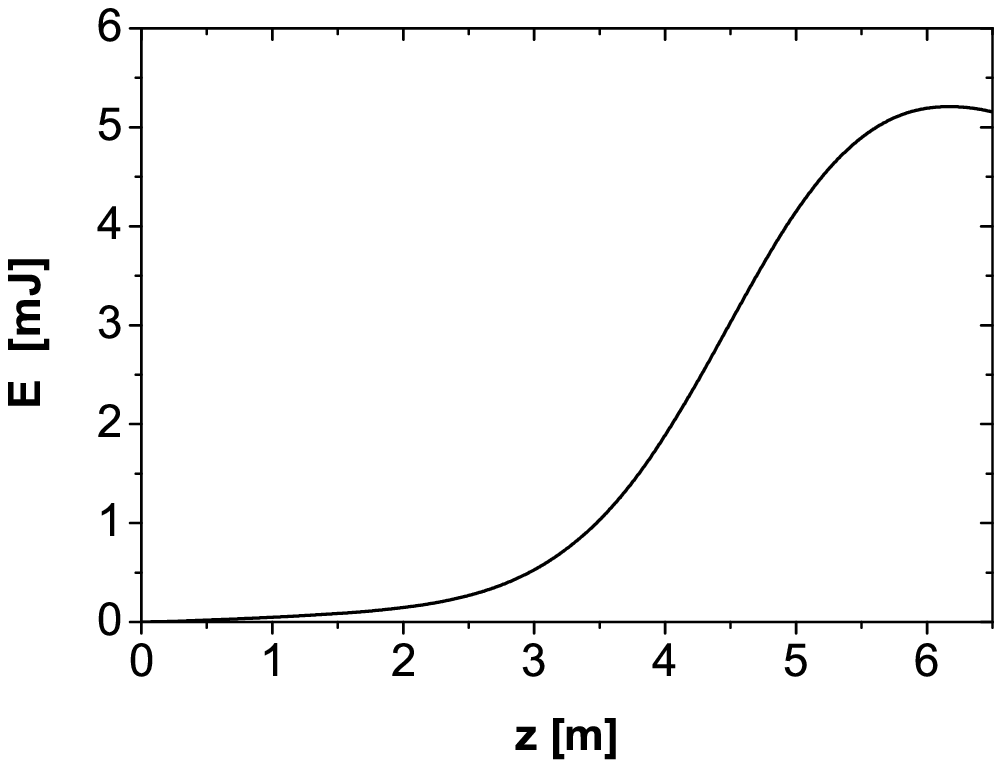,width=0.5\textwidth}

\vspace*{-63mm}

\hspace*{0.5\textwidth}
\epsfig{file=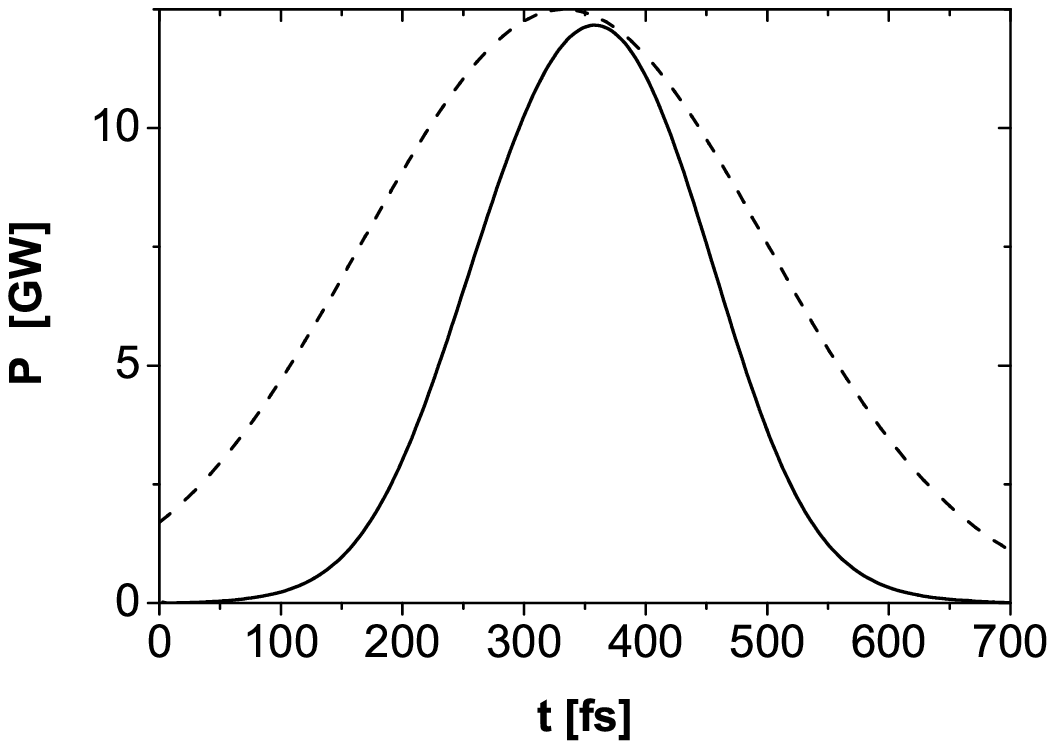,width=0.5\textwidth}

\caption{
Radiation energy in the optical pulse versus undulator length
(left plot) and time structure of the radiation pulse at the exit of the
optical radiator (right plot) in short wavelength mode for 400~nm
wavelength.
The dashed line shows the bunch profile.
}
\label{fig:r400-prj}
\end{figure}

\begin{figure}[tb]

\epsfig{file=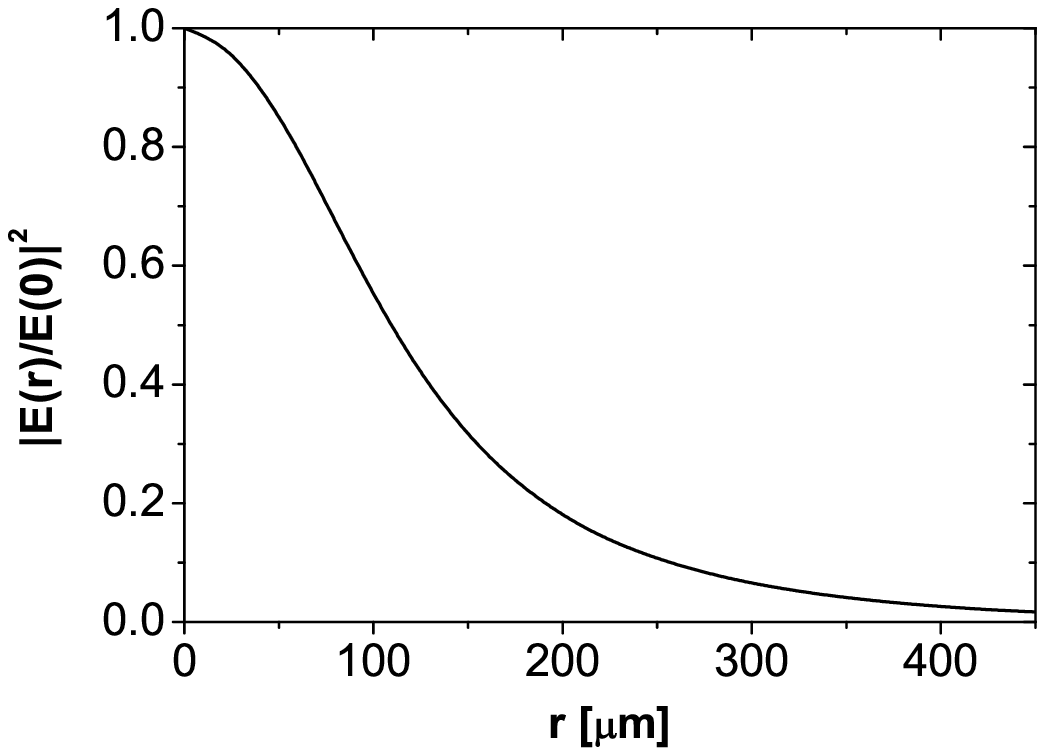,width=0.5\textwidth}

\vspace*{-63mm}

\hspace*{0.5\textwidth}
\epsfig{file=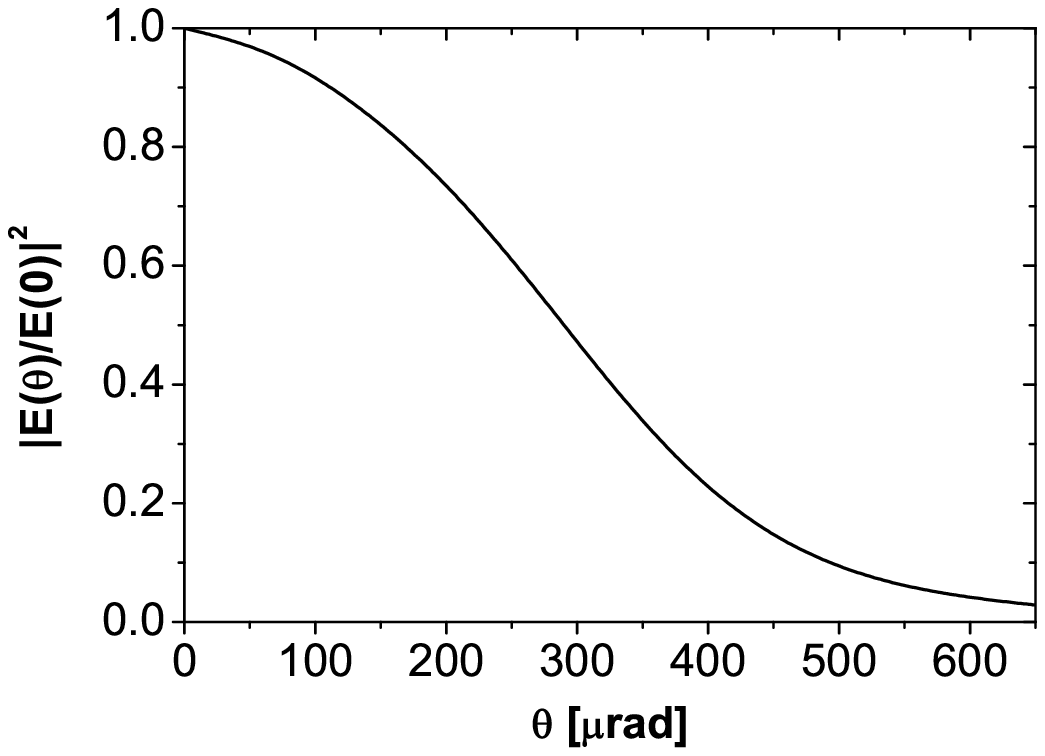,width=0.5\textwidth}

\caption{
Radial distribution of the optical pulse intensity at the undulator exit
(left plot) and intensity distribution in the far zone (right plot) in short
wavelength mode for 400~nm wavelength.
}
\label{fig:nf-prj-400}
\end{figure}

\begin{figure}[p]
\begin{center}
\epsfig{file=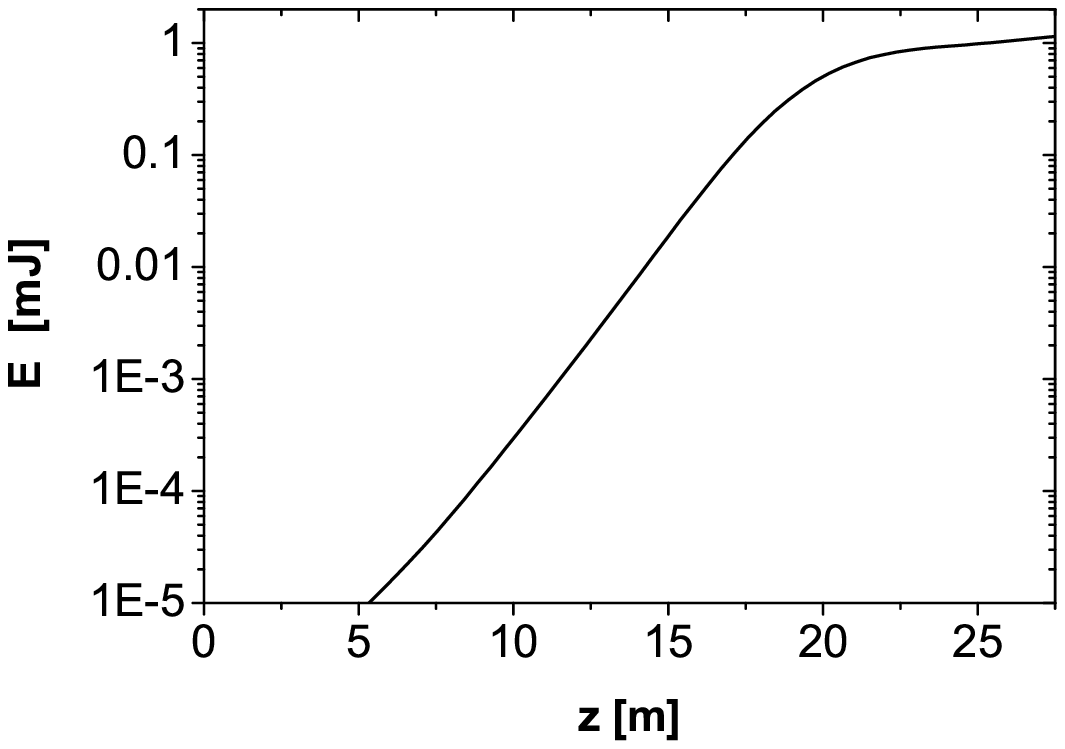,width=0.6\textwidth}
\end{center}
\caption{
Radiation energy of the X-ray SASE pulse versus undulator length in the
short wavelength mode for a
radiation wavelength of 6.4~nm.
}
\label{fig:pz6-prj}


\epsfig{file=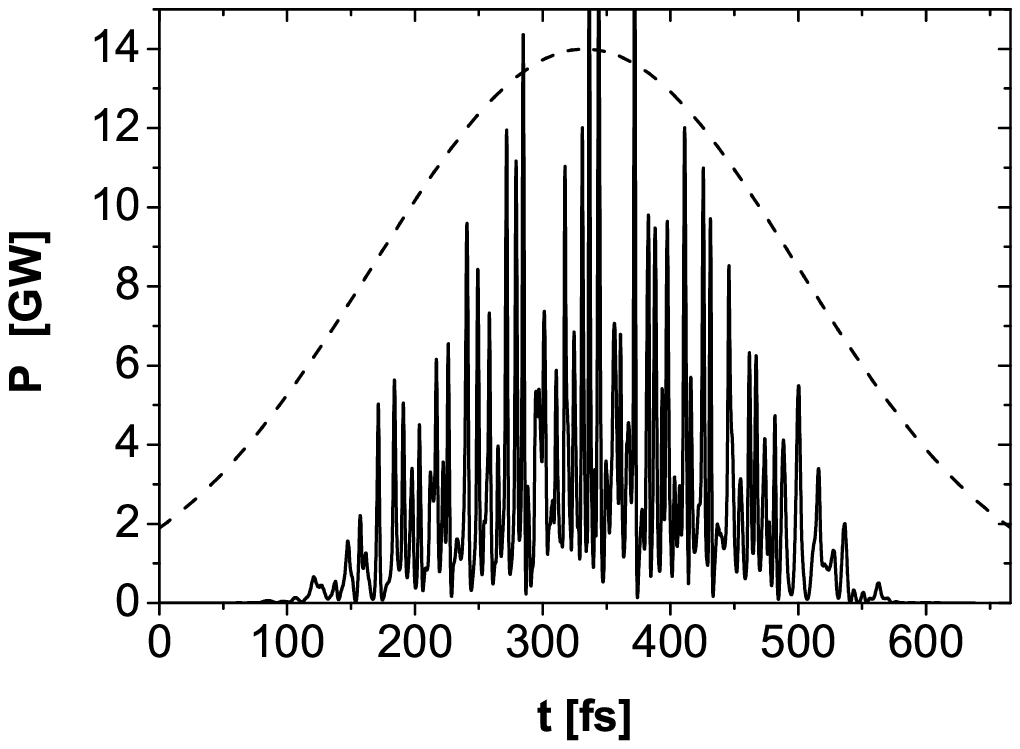,width=0.5\textwidth}

\vspace*{-63mm}

\hspace*{0.5\textwidth}
\epsfig{file=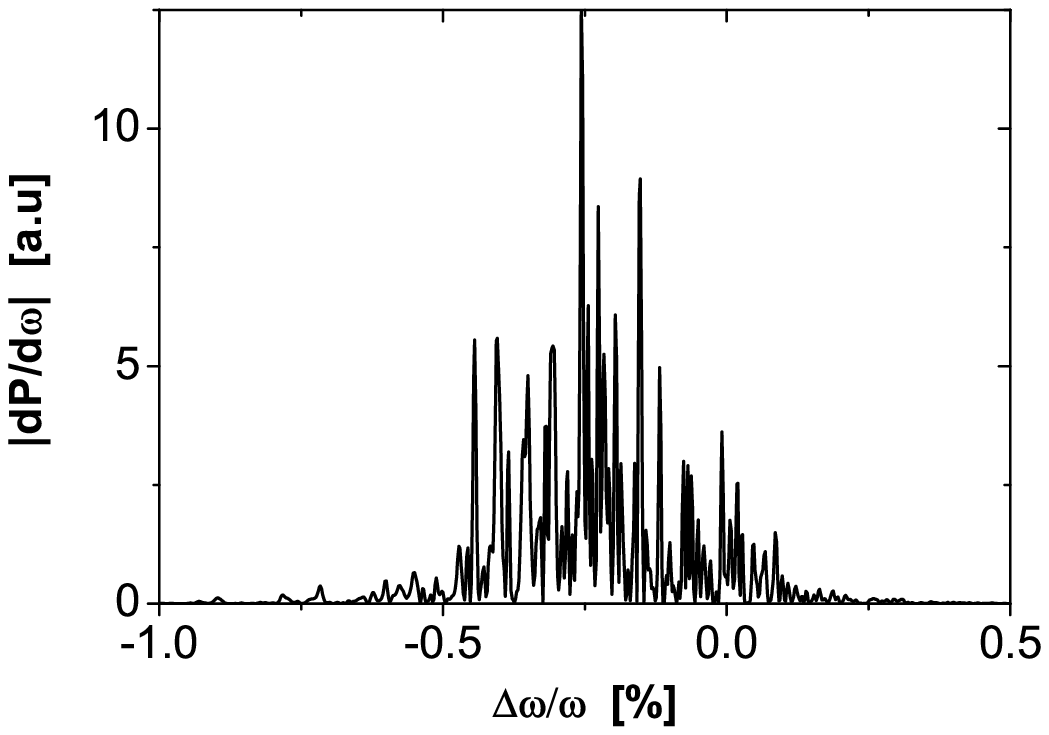,width=0.5\textwidth}

\caption{
Time (left plot) and spectral (right plot) structure of the X-ray
SASE pulse at the exit of the undulator in the short wavelength mode for a
radiation wavelength of
6.4~nm.
The dashed line shows the bunch profile.
}
\label{fig:ts6-prj}


\epsfig{file=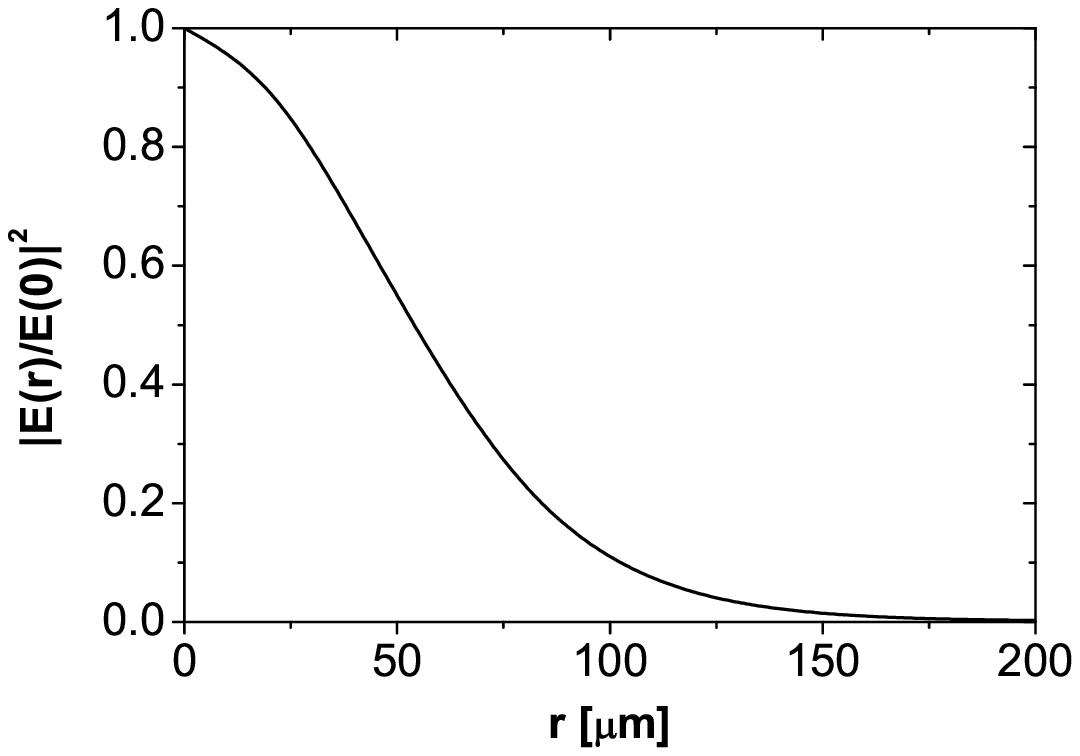,width=0.5\textwidth}

\vspace*{-63mm}

\hspace*{0.5\textwidth}
\epsfig{file=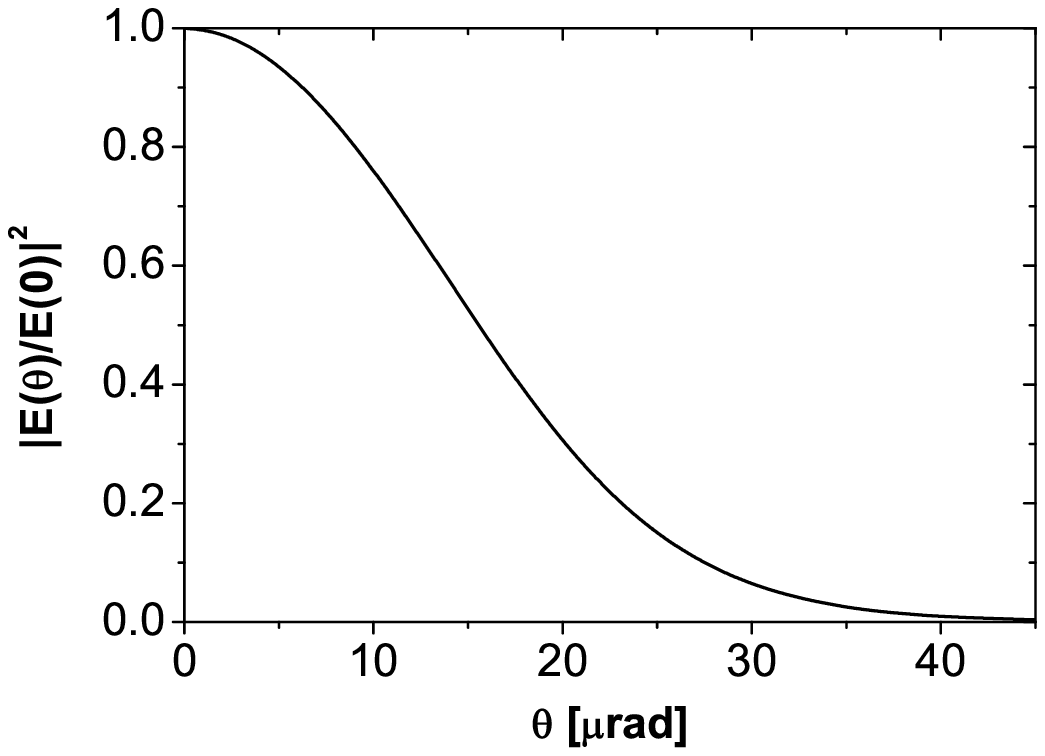,width=0.5\textwidth}

\caption{
Radial distribution of the X-ray SASE pulse intensity at the undulator exit
(left plot) and its intensity distribution in the far zone (right plot) for
the short wavelength mode at a.
radiation wavelength of 6.4~nm.
}
\label{fig:nf-prj-6}
\end{figure}

The operation of the two-color scheme is the same as it was described
above.  The electron beam is modulated in the optical modulator (see
Fig.~\ref{fig:mod-fem}). Then it passes through the X-ray undulator and
produces GW-level X-ray radiation pulses.  Upon leaving the X-ray
undulator the beam is directed to the optical radiator and produces
GW-level optical pulses. The properties of the optical radiation are
illustrated with Figs.~\ref{fig:r400-prj} and \ref{fig:nf-prj-400}.
The  energy of the optical pulse exceeds the mJ level, and the peak
radiation power exceeds 10~GW. The FWHM pulse duration is about 150~fs.

The properties of the X-ray pulse are illustrated with
Figs.~\ref{fig:pz6-prj}, \ref{fig:ts6-prj}, and \ref{fig:nf-prj-6}.  It
is seen that the properties of the X-ray radiation are the same as for
an unmodulated electron beam \cite{ttf-phase2}.  Thus, we can state
that the two-color facility does not interfere with the main modes of
FEL operation.

\subsection{Pulse separation}

We demonstrated that the proposed two-color facility is capable of
producing GW-level optical and X-ray pulses which are precisely
synchronized at a femtosecond level. These pulses overlap not only
longitudinally, but also transversely if the X-ray undulator and the
optical radiator are in line. In this case the two beams can be
separated for pump-probe experiments by making use of their rather
different divergence (see, e.g.  Figs.~\ref{fig:nf-fem-400} and
\ref{fig:nf-fem-30}). At some distance from the source the optical beam
size will be much larger than that of the X-ray beam such that a mirror
with a hole can be used to separate the beams.  Another possibility is
to tilt the undulator axes by about a mrad.  In this case the optical
and X-ray pulses are pointing in slightly different directions and can
be delivered to the sample via different beamlines. The optical transport line to the sample has to include a variable delay which allows precise tuning over a range of several picoseconds including zero crossing. This is not trivial since only very small deflection angles are possible for the X-ray beam, and there are other geometrical constraints coming from the existing FEL beam distribution system. This needs further work and is beyond the scope of the present paper.

\section{Conclusion}

\begin{table}[b]
\caption{Properties of the radiation pulses of the two-color pump-probe
facility}
\medskip
\begin{tabular}{l c c c }
\hline
Parameter &
Units & Femtosecond & Short wavelength \\
& & mode & mode \\
\hline \\
\underline{Optical pulse} \\
\hspace*{10pt} Wavelength             & nm & \multicolumn{2}{c}{200-900} \\
\hspace*{10pt} Pulse energy           & mJ & 0.1--0.5 & 1--5 \\
\hspace*{10pt} Pulse duration (FWHM)  & fs & 30-100   & 150 \\
\hspace*{10pt} Peak power             & GW & 2        & 5--15 \\

\hspace*{10pt} Spectrum width         &   &
\multicolumn{2}{c}{Transform-limited} \\
\hspace*{10pt} Spot size (FWHM)       & $\mu $m & 150--200 & 80--120 \\
\hspace*{10pt} Angular divergence$^*$ (FWHM) & $\mu $rad & 100--500 &
150--700 \\
\hspace*{10pt} Repetition rate        & Hz & \multicolumn{2}{c}{10 $(10^4)$}
\\
\underline{X-ray SASE pulse} \\
\hspace*{10pt} Wavelength             & nm & 30--120  & 6--30 \\
\hspace*{10pt} Pulse energy           & mJ & 0.1 &  1 \\
\hspace*{10pt} Pulse duration (FWHM)  & fs & 30-100   & 150 \\
\hspace*{10pt} Peak power             & GW & 2        & 2 \\
\hspace*{10pt} Spectrum width         & \%  & 0.4--0.6 & 0.3--0.6\\
\hspace*{10pt} Spot size (FWHM)       & $\mu $m & 350--1400 & 140--210 \\
\hspace*{10pt} Angular divergence (FWHM) & $\mu $rad & 40--150 & 20--70 \\
\hspace*{10pt} Repetition rate        & Hz & \multicolumn{2}{c}{10 $(10^4)$}
\\
\hline
\end{tabular}

$^*$Diffraction limited

\label{tab:pulse-properties}
\end{table}

A novel two-color FEL amplifier for pump-probe experiments has been described combining sub-100~fs
optical and X-ray pulses. The properties of the radiation pulses are
summarized in Table~\ref{tab:pulse-properties}.  The proposed facility
has unique features: Both pulses have very high peak power in the  GW
range.  The wavelengths of both radiation sources are continuously
tunable in a wide range:  $200-900$~nm for the optical pulses, and
$6-120$~nm for the X-ray pulses. Both pulses have diffraction limited
angular divergence.  The spectral width of the optical pulse is
transform limited. Finally and most important, optical and X-ray pulses
are precisely synchronized at a femtosecond level, since they both are
produced by the same electron bunch, and there are no reasons for any
time jitter between the pulses.  Based on these unique features a
pump-probe facility could be built with unique possibilities for studying time dependent processes on the time scale of chemical reactions.  It is worth to mention that the Nobel prize in chemistry in
1999 was awarded to A.~Zewail for pump-probe experiments using a
quantum laser (40~fs pulse duration) operating in the visible range. The combination of visible light and X-rays would add the new dimensions of element specificity and direct structural information.

A two-color FEL amplifier could be realized at the TESLA Test Facility rather quickly  and with moderate cost expenses for the required components, i.e. a seed laser, two optical undulators, and a
dispersion section The tunable-gap
optical undulators would be similar to insertion devices used at DORIS.
Initially one could use a commercially available long (nanosecond) pulse dye laser
with a repetition rate of 10~Hz. This could later be replaced by an OPA
laser system similar to the one currently under development at TTF \cite{pp} in order 
to operate the system at the full repetition rate of the LINAC.

\section*{Acknowledgments}

We thank R.~Brinkmann, J.~Krzywinski, E.A.~Matyushevskiy, J.~Rossbach,
and M.~Tischer for many useful discussions.  We thank C.~Pagani,
J.R.~Schneider, D.~Trines, and A.~Wagner for interest in this work.

\end{document}